\documentclass[journal,10pt,twocolumn]{IEEEtran}
\usepackage{url}
\usepackage{rotating}
\usepackage{cite}
\usepackage{amsthm}
\usepackage{amssymb}
\usepackage{amsmath}
\usepackage{amsfonts}   
\usepackage{amssymb}    
\usepackage{float}
\usepackage{color}
\usepackage{graphicx}
\usepackage{epsfig}
\usepackage{epstopdf}
\usepackage{pbox}
\usepackage{rotating}
\usepackage[singlespacing]{setspace}
\usepackage{multirow}
\newcommand{\figref}[1]{Fig.~\ref{#1}}
\theoremstyle{plain}

\theoremstyle{plain}

\theoremstyle{plain}
\newtheorem{lemmacounter}{Theorem}
\newtheorem{lemma}[lemmacounter]{Lemma}
\theoremstyle{plain}
\newtheorem{defcounter}{Theorem}
\newtheorem{definition}[defcounter]{Definition}
\theoremstyle{plain}

\newtheorem{special case}[defcounter]{Special Case}
\usepackage{url}
\usepackage{rotating}
\usepackage{cite}
\usepackage{amsthm}
\usepackage{amssymb}
\usepackage{amsmath}
\usepackage{amsfonts}   
\usepackage{amssymb}    
\usepackage{float}
\usepackage{color}
\usepackage{graphicx}
\usepackage{epsfig}
\usepackage{epstopdf}
\usepackage{pbox}
\usepackage{rotating}
\usepackage[singlespacing]{setspace}
\usepackage{multirow}
\begin{document}

\title{Cellular Downlink Performance with Base Station Sleeping, User Association, and Scheduling\\[10pt] \thanks{$^\dag$Hina Tabassum, Uzma Siddique, and Ekram Hossain are with the  Department of Electrical and Computer Engineering, University of Manitoba,  Canada (emails: \{hina.tabassum, uzma.siddique, ekram.hossain\}@umanitoba.ca). Md. Jahangir Hossain is with the School of Engineering, University of British Columbia (Okanagan campus), Kelowna, BC, Canada (email: jahangir.hossain@ubc.ca).}
\author{\IEEEauthorblockN{Hina Tabassum, Uzma Siddique, Ekram Hossain, and Md. Jahangir Hossain}}
}

\maketitle
\IEEEpeerreviewmaketitle

\begin{abstract}
Base station (BS) sleeping has emerged as a viable solution to enhance the overall network energy efficiency by inactivating the underutilized BSs. However,  it affects the performance of users in sleeping cells  depending on their BS association criteria, their channel conditions towards the active BSs, and scheduling criteria and traffic loads at the active BSs. This paper characterizes the performance of  cellular systems with BS sleeping  by developing a  systematic framework to derive the spectral efficiency and outage probability of downlink transmission to the sleeping cell users taking into account the aforementioned factors. In this context, we develop a  user association scheme in which a typical user in a sleeping cell selects a BS with \textbf{M}aximum best-case \textbf{M}ean channel \textbf{A}ccess \textbf{P}robability (MMAP) which is calculated by all active BSs based on their existing traffic loads. We consider both  greedy and round-robin schemes at active BSs for scheduling users in a channel.  Once the association is performed, the exact access probability for a typical sleeping cell user and the statistics of its received signal and interference powers are derived to evaluate the spectral and energy efficiencies of transmission. For the sleeping cell users, we also consider the conventional \textbf{M}aximum \textbf{R}eceived \textbf{S}ignal \textbf{P}ower (MRSP)-based user association  scheme along with greedy and round-robin schemes at the BSs. 
The impact of cell-zooming is incorporated in the derivations to analyze its feasibility in reducing the coverage holes created by BS sleeping.  
Numerical results show the trade-offs between spectral efficiency and  energy efficiency in various network scenarios. The accuracy of the analysis is verified through Monte-Carlo simulations. 
\end{abstract}

\section{Introduction}
Energy efficiency has become one of the highly desirable characteristics of the future wireless cellular networks. 
Conventional energy saving methods aim to reduce the transmit power of the base stations (BSs); however, recent studies have shown that 60-80\%  of the energy consumed per BS is static (i.e., due to battery backup, internal processing, air conditioning, etc.) and independent of traffic load \cite{MAMarsan}. Moreover, it has also been reported that  BSs are largely underutilized, i.e., the fraction of time during which  traffic load remains below 10\% is estimated to be 30\% in weekdays  and 45\% at weekends \cite{6}.  Thus, the traffic fluctuations provide a significant margin to improve the overall network energy efficiency by allowing the underutilized BSs to sleep~\cite{Oliver}. Nonetheless,  maintaining the coverage and spectral efficiency  of transmission for the users in sleeping cells is an important issue that needs careful investigation. In this regard, techniques such as  cell-breathing (or cell-zooming) \cite{5} and coordinated multi-point transmissions (COMP)  \cite{han2011energy} are  envisioned to provide coverage to the sleeping cell users  and establish a  balance between the overall network performance and energy savings.

\subsection{Related Work}
The performance of BS sleeping has been analyzed in 
several algorithmic and simulation-centric studies with different scenarios, assumptions, and network parameters (see \cite{MAMarsan,bhaumik2010breathe,5, han2011energy,14,cao1} and references therein). In \cite{MAMarsan}, optimal energy savings are calculated as a function of the daily traffic pattern while ignoring the wireless link factors. In \cite{bhaumik2010breathe}, network energy consumption is minimized by adjusting the cell sizes when BS sleeping is performed. Nevertheless, the transmit power required by the active BSs to provide coverage to  sleeping cell users, i.e., effect of cell-zooming is  ignored in the energy efficiency analysis \cite{MAMarsan,bhaumik2010breathe}.  Other interesting studies optimize network utility functions, e.g., minimize energy expenditure by optimizing the BS activity with traffic load constraints \cite{14} and minimize a weighted combination of power consumption  and delay to optimize the service rate \cite{cao1}.

Recently, few analytical frameworks have considered to derive the optimal density of sleeping BSs~\cite{Soh2013,powersleepmode2,cao}, proposed distributed sleeping mechanisms  considering cognitive small cells \cite{wilde}, and developed different BS switching-off schemes \cite{distanceaware, han}. The development of such  frameworks is highly desirable to capture the joint interplay among the  network design parameters while extracting in-depth  theoretical insights behind  various observations and performance trends.   
A power minimization problem with coverage constraint for a randomly selected sleeping cell user is formulated in \cite{Soh2013,powersleepmode2} such that the BS density is optimized considering random  and traffic load-based sleeping. 
To avoid coverage holes, power of all active BSs is assumed to increase equally regardless of the channel conditions of the sleeping cell users.
Due to the complexity of closed-form solution,  
an optimal value of BS density is computed numerically. 
In \cite{cao}, a BS density minimization problem is formulated with coverage constraints of users in  sleeping cells. However, the  expression for coverage includes two improper integrals and an infinite summation. Therefore, the authors  opt to  derive upper and lower bounds on the optimal BS density.

A distance-aware BS switching algorithm is presented in \cite{distanceaware} that recommends to switch-off a BS with maximum average distance from its own and neighboring cell users. In \cite{han}  four heuristic-based BS switching-off patterns are proposed and  the coordination among the BSs is exploited  to provide coverage to the sleeping cell users. 

\subsection{Motivations and Contributions}
The aforementioned studies mainly focus on improving the energy efficiency of cellular networks  by either optimizing the density of active BSs with certain coverage constraint for the sleeping cell users or proposing different switching-off schemes/patterns and analyzing their relative performance gains. 
The impact of  channel availability on the performance of sleeping cell users, which depends on the scheduling criteria and traffic loads at the active BSs, has been largely ignored in the previous studies.  For example, under same traffic load conditions, if an active BS uses greedy scheduling on a given transmission channel, the probability of channel availability to a sleeping cell user is significantly low compared to that for round-robin scheduling. The multi-user scheduling criteria used at the active BSs have a direct consequence on the channel access probability and the achieved spectral efficiency of transmission to the sleeping cell users as well as the spectral and energy efficiencies of the entire network.

In the above context, given a number of active BSs or switching-off pattern, this paper develops a rigorous analytical framework to derive the spectral efficiency as well as the outage probability of a typical user in the underutilized sleeping cells. We assume that all BSs share the same transmission channels and perform scheduling on each transmission channel depending on the network objective (e.g., throughput maximization  or fairness maximization). The coverage  to sleeping cell users is  ensured by exploiting the BS zooming capabilities\footnote{From the system design perspective  both cell-zooming and BS cooperation are equally important  techniques for coverage expansion. However, the comparison between these two techniques is beyond the scope of this paper.}. 

The major contributions of the paper can be summarized as follows:

\begin{enumerate}
\item This paper develops a systematic framework to
characterize the performance of  cellular systems with BS sleeping. The framework  captures the impact of  {channel availability} on the performance of a typical sleeping cell user which depends on the channel conditions of the sleeping cell user towards the active BSs and traffic loads at the active BSs (or the scheduling criteria used at the active BSs).  Considering both greedy and round-robin scheduling schemes at the active BSs and their existing traffic loads, for a typical sleeping cell user, we derive the {\em best-case mean channel access probability} (i.e., the highest probability of obtaining a channel allocation, which is spatially averaged over the cellular region) corresponding to the different active BSs.

\item Based on the derived best-case channel access probabilities, we develop a  user association scheme in which a typical user in a sleeping cell selects a BS (among all active BSs) with the maximum best-case mean channel access probability (MMAP). Once the association is performed, the exact access probability for a typical sleeping cell user and the statistics of its received signal and interference powers are derived. We compare this with conventional maximum received signal power (MRSP)-based user association  scheme. We show that MRSP-based association may not always be the best choice and that the channel access probabilities from active BSs
serve as a useful user association metric.  Design guidelines are also provided for simple hybrid schemes that can be implemented to overcome the drawbacks of MMAP and MRSP-based user association schemes.

\item The outage probability and spectral efficiency of a typical sleeping cell user are then evaluated. The impact of cell-zooming on reducing the
coverage holes is analyzed. To reduce the complexity of evaluation based on the exact analytical expressions, we also present an approximate approach to characterize the performance of a typical sleeping cell user in all considered scenarios. 

\item Finally, for various network scenarios, the trade-offs between spectral efficiency and energy efficiency are captured  and the accuracy of the analysis is validated through Monte-Carlo simulations.  The feasibility of BS sleeping is analyzed by quantifying the transmit power required at a given active BS to maintain the throughput  of a sleeping cell user as it was before (i.e., when its corresponding BS was active).   Numerical results demonstrate that the required zooming power  depends significantly on the scheduling criteria used at the active BSs and the association scheme used by the sleeping cell users. 

\end{enumerate}

\subsection{Organization and Notations}
The  remainder of this paper is organized as follows.  Section II details the network and
channel models, MMAP and MRSP-based association schemes for users in the sleeping cells, distance distributions, power consumption model, and also outlines the methodology of analysis. In Section~III, we derive the best-case mean channel access probability of a sleeping cell user considering  greedy and round-robin  schemes at the active BSs. Based on this, in Section~IV, we derive the association probabilities for a sleeping cell user with  different active BSs considering both MMAP and MRSP-based associations. Also, derive expressions for the exact channel access probability and  received signal and interference powers for the sleeping cell user depending on the user association schemes. Next, in Section V, we derive coverage and spectral efficiency expressions for all considered cases. Section~VI presents  numerical  results followed by concluding remarks in Section~VII.

\vspace{1 mm}
\noindent\textbf{Notations}: 
$\mathrm{Gamma}(\kappa_{(\cdot)},\Theta_{(\cdot)})$ represents a Gamma distribution with shape parameter $\kappa$, scale parameter $\Theta$ and $(\cdot)$ displays the name of the random variable (RV). $\mathcal{K}_G (m_{c_{(\cdot)}},m_{s_{(\cdot)}},\Omega_{(\cdot)})$ represents the generalized-$\mathcal{K}$ distribution with  fading parameter $m_c$, shadowing parameter $m_s$ and average power $\Omega$. $\Gamma(a)=\int_0^\infty x^{a-1} e^{-x} dx$ represents the Gamma function, ${\Gamma}_u (a;b)=\int_b^\infty x^{a-1} e^{-x} dx$ denotes the upper incomplete Gamma function, ${\Gamma}_l(a;b)=\int_0^b x^{a-1} e^{-x} dx$ denotes the lower incomplete Gamma function and ${\Gamma}(a;b_1;b_2)=\Gamma_u(a;b_1)-\Gamma_u(a;b_2)=\int_{b_1}^{b_2} x^{a-1} e^{-x} dx$ denotes the generalized Gamma function \cite{book}. $_2F_1[\cdot,\cdot,\cdot,\cdot]$ denotes the Gauss's hypergeometric function.
$\mathrm{Pr}(A)$ denotes the probability of event $A$. $f(\cdot)$, $F(\cdot)$, and $\mathcal{M}(\cdot)$ denote the probability density function (PDF), cumulative distribution function (CDF), and moment generating function (MGF), respectively. Finally, $\mathbb{U}(\cdot)$, $\delta(\cdot)$, and $\mathbb{E}[\cdot]$ denote the unit step function, the Dirac-delta function, and   the expectation operator, respectively. A list of the main notations  and their definitions is given in Table~I.

\small
\begin{table*}[!ht]
\caption{Summary of the main notations and their definitions}
     \begin{tabular}{ | l  | p{16cm} |}
     \hline
Variable & Definition
\\
\hline
$\gamma_{k}$ & Channel between the active BS $k$ and its local users
\\
${r}_{k}$ & Distance between the user  in active cell $k$  and BS $k$ 
\\
$X_{jk}$ & Channel between the active BS $k$ and a typical user of sleeping cell $j$
\\
$D_{jk}$ & Distance between the BS of sleeping cell user of interest and active BS $k$   
\\
$\tilde{r}_{jk}$ & Distance between the  sleeping cell user of interest in BS $j$ and active BS $k$  
\\
$\mathcal{F}_{k^*}$& Set of sleeping cells whose users are associated to $k^*$
\\
$\mathcal{A}$ & Set of active BSs
\\
$\mathcal{S}$ & Set of sleeping BSs
\\
$\mathcal{Q}$ & Outage threshold
\\
$\tilde{r}_{w,y}$ & Discrete distance between the  sleeping cell user of interest located on polar coordinate $(r_y, \theta_w)$ in BS $j$ and active BS $k$, $r_y \in \{r_1,r_2,\cdots,r_{\mathcal{Y}}\}$, $\theta_w \in \{\theta_1,\theta_2,\cdots,\theta_{\mathcal{W}}\}$ 
\\
$\tilde{p}_{jk}^{(\cdot)}$ & Best-case channel access probability of a user in sleeping cell $j$ from an active BS $k$;  ($\cdot$) denotes $\mathrm{GR}$  and $\mathrm{RR}$ for greedy and round-robin scheduling, respectively  
\\
$\hat{p}_{jk}^{(\cdot)}$ & Association probability of a user in sleeping cell $j$ from an active BS $k$; 
($\cdot$) denotes $\mathrm{MMAP}$  and $\mathrm{MRSP}$ for MMAP and MRSP-based user association schemes, respectively
\\
$S_{jk^*}$ & Received signal power of a typical sleeping cell user in cell $j$ when associated with BS $k^*$
\\
$I^{\mathrm{cum}}_{jk^*}$ & Cumulative interference at a typical sleeping cell user in cell $j$ when associated with BS $k^*$
\\
\hline
\end{tabular}
\end{table*}
\normalsize

\section{System Model and Assumptions}

In this section, we  describe the considered cellular network model, user association schemes, distance distributions,  power consumption model, and outline the methodology of analysis.

\subsection{Network Model}
 
We consider a downlink network of $L$ circular macrocells\footnote{The regular cellular deployment may not exist in reality; however,  the deployment of macrocell BSs is always well planned. Thus, such deployments are used to develop tractable frameworks in order to analyze the system performance in practical scenarios of interest\cite{lte}.}, each of radius $R$ and $U_l, \:\:\forall l=1,2,\cdots,L,$ uniformly distributed users. 
The frequency reuse factor is assumed to be unity, i.e., each transmission channel is reused in all cells. Each BS selects a user on a given transmission channel or subcarrier\footnote{Typically for resource allocation type 0,  a resource block (RB) composed of several consecutive subcarriers is allocated to a user instead of a  subcarrier. Nonetheless, since all subcarriers in a RB allocated to a user experience nearly the same channel conditions \cite{rb1,rb2}, their corresponding signal, interference powers, and spectral efficiency remain nearly the same and can be summed up to provide  spectral efficiency of a user per RB rather than per subcarrier.} in orthogonal frequency division multiple access (OFDMA) networks considering a predefined scheduling criterion.
Multi-user scheduling on each transmission channel is required to  assign resources such that the desired network objective (e.g., maximize throughput, maximize fairness) can be achieved  \cite{TWChina,lte,geof,mobility}. In this regard, various scheduling schemes are available in the literature. For example,  greedy scheduling maximizes the throughput on each channel by selecting a user with best received signal power. On the other hand, the well-known round-robin scheduling \cite{Soh2013} maximizes the system fairness by selecting a user arbitrarily regardless of its channel conditions. Proportional fair \cite{TWChina} and opportunistic round-robin \cite{orr} are two other popular scheduling schemes that lie in between the two extremes in terms of throughput and fairness trade-off. In this paper, we  focus on the greedy and round-robin schemes for scheduling users on a given transmission channel.   Our main motivation for this is to investigate the impacts of two extreme scheduling schemes (one of which maximizes cell throughput while the other maximizes  fairness among users in a cell)  on the spectral and  energy efficiencies of transmission to sleeping cell users.  

\begin{figure*}[t]
\begin{center}
\includegraphics[width=4.75in]{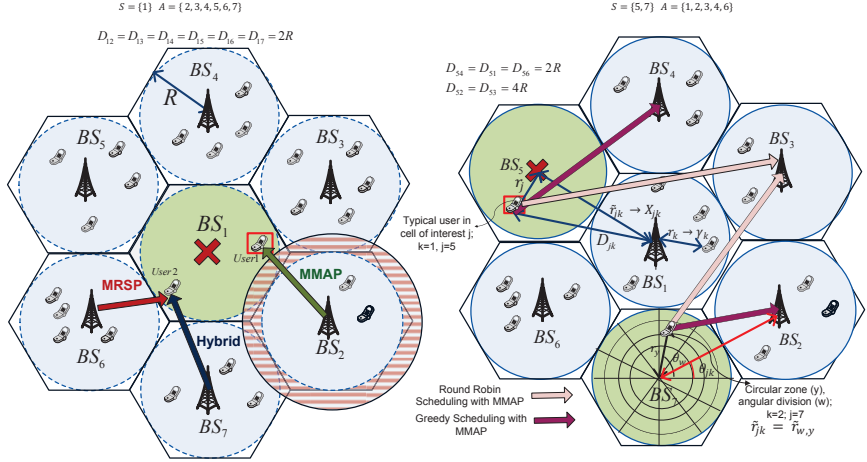}
\caption [c]{
Graphical illustration of the first tier of circular macrocells with different BSs in sleeping and zooming modes. The working mechanisms of MMAP and  MRSP-based  user associations are illustrated in the first part of the figure. Graphical demonstration of the distances used in exact and approximate approaches (to be presented in Section~III) are labeled in the second part. The MMAP-based association for greedy and round-robin scheduling schemes in different sleeping cells is illustrated in the second part of the figure.
}
\label{Multi-Tier}
\end{center}
\end{figure*}

We assume that accurate channel state information (CSI) available at the BS to implement greedy (i.e., opportunistic) scheduling on a transmission channel\footnote{This  is more realistic for  low-mobility users. For high mobility users, the corresponding CSIs may become inaccurate and the gain of greedy scheduling  may not be significant compared to that of round-robin scheduling.}.
The fraction of sleeping BSs is represented by $q$, and consequently, the number of sleeping and active BSs can be given as $N_{\mathrm{sleep}}=L q$ and $N_{\mathrm{active}}=L(1-q)$, respectively.
The sets of sleeping and active BSs are denoted by $\mathcal{S}$ and $\mathcal{A}$, respectively.  
A graphical illustration of the considered system model with first tier of circular macrocells is shown in \figref{Multi-Tier}. The sets of sleeping and active BSs are illustrated graphically for two different switching-off patterns and a typical  user of interest is shown to be located in sleeping cell $j \in \mathcal{S}$.  

The number of inactive cells $N_{\mathrm{sleep}}$ and their locations can be selected  either by using different switching-off patterns \cite{han} or by specifying a certain number of users $U_{\mathrm{th}}$ that represents a low-load condition, i.e., a BS remains inactive if the number of users in the corresponding cell is less than or equal to $U_{\mathrm{th}}$. The active BSs provide coverage to the users in sleeping cells by increasing their transmit power. The users in the active BSs are assumed to remain associated with their closest BS.

\subsection{User Association Schemes}
We consider two user association schemes, namely, (i) \textbf{M}aximum \textbf{M}ean channel \textbf{A}ccess \textbf{P}robability (MMAP) and (ii) \textbf{M}aximum \textbf{R}eceived \textbf{S}ignal \textbf{P}ower (MRSP) schemes, for sleeping cell users. MMAP represents a user association scheme in which a typical user of interest in sleeping cell $j \in \mathcal{S}$ associates to the BS $k \in \mathcal{A}$ from which it has the maximum  chance of obtaining the channel allocation, i.e., a BS which offers the maximum  access probability.  We assume that the mean channel access probability of a sleeping cell user from an active BS $k$ is broadcast by  BS $k$ immediately after some BSs in the network become inactive and when any potential sleeping cell user asserts a request for this information from  BS $k$.
Thus, the MMAP-based association  is a network-assisted user association scheme in which a sleeping cell user selects the BS given the  mean channel access probabilities provided by different active BSs. 

Nonetheless, since more than one sleeping cell users can associate to this BS, this  mean channel access probability depends on the original number of users associated to BS $k$ and may not reflect the exact channel access  probability that a sleeping cell user will experience after associating to BS $k$. We therefore refer to this probability  as the \emph{best-case mean channel access probability}. Note that the MMAP-based user association depends on the scheduling scheme used at the base station and hence depends on the traffic loads at the active cells and the channel conditions of the sleeping cell users towards the active BSs.

MRSP represents the conventional  user association scheme where a user of interest selects a BS from which it experiences the  maximum instantaneous received signal power. This criterion requires instantaneous channel information at a user from all active BSs and  is independent of  channel access probabilities and thus the traffic load conditions corresponding to the different active BSs.
 
For illustration, consider a scenario shown in first part of \figref{Multi-Tier}, where ${\rm user~1}$ in the coverage area of BS$_1$ (which is in sleep mode) may associate to BS$_2$ or BS$_3$ with MMAP and MRSP-based associations, respectively. Similarly, ${\rm user~2}$ will associate to BS$_2$ or BS$_6$ with MMAP and MRSP-based association, respectively. However, since BS$_2$ is quite far from ${\rm user~2}$,  a better option for ${\rm user~2}$ might be to associate with BS$_7$ rather than with BS$_2$ based on a {\em hybrid} user association scheme, which takes into account both traffic load and channel condition. Note that, in this paper we primarily focus on MMAP and MRSP-based association schemes. Design guidelines for  simple hybrid schemes that overcome the drawbacks of both MMAP and MRSP user association schemes will be  provided in Section~V.

\subsection{Power Consumption Model}

The total network power consumption before BS sleeping is modeled as~\cite{arnold2010power}:
$
{P}_{\mathrm{tot}} = L (\Delta_{\mathrm{dyn}}P_t  +P_{\mathrm{static}}), 
$
where $P_{\mathrm{static}}=P_{\mathrm{SP}}(1+C_c)(1+C_{PSBB})$ represents the static power consumption
due to signal processing overhead ($P_{\mathrm{SP}}$), battery backup and power supply losses ($C_{\mathrm{PSBB}}$), and  cooling losses ($C_c$), $\Delta_{\mathrm{dyn}}=(1+C_c)(1+C_{PSBB})/\eta_{\mathrm{PA}}$ represents 
the slope of load-dependent power consumption of the battery backup, cooling, and power supply losses as well as power amplifier efficiency~\cite{powersleepmode1,arnold2010power}.  When 
a fraction of $q$ BSs are allowed to sleep, the reduced power consumption can be given as
\begin{align}\label{p}
\tilde{P}_{\mathrm{tot}}&= N_{\mathrm{active}} (\Delta_{\mathrm{dyn}} P_t +P_{\mathrm{static}})  +P_{\mathrm{sleep}} N_{\mathrm{sleep}}, 
\nonumber\\&=(1-q) {P}_{\mathrm{tot}}+P_{\mathrm{sleep}} N_{\mathrm{sleep}},
\end{align} 
where $\mathrm{P}_{\mathrm{sleep}}$ is the BS power  consumption while sleeping; however, it is negligibly small compared to the static power consumption \cite{powersleepmode1,powersleepmode2}. 
To avoid coverage holes, the amount of increase in transmit power required per channel is given by $\tilde{P}_t=\alpha P_t$, where $\alpha$ is a scaling factor, i.e., $\alpha > 1$ when  cell-zooming is performed and $\alpha = 1$ for no  cell-zooming scenario. Note that a BS will zoom if any of the associated sleeping cell users is selected for transmission. The net power consumption after cell-zooming can then be derived as: 
$
\bar{P}_{\mathrm{tot}}=  
N_{\mathrm{zoom}}  {P}_t \Delta_{\mathrm{dyn}} (\alpha-1) +\tilde{P}_{\mathrm{tot}},
$
where ${N_{\mathrm{zoom}}}$ represents the number of BSs in zooming mode.

\subsection{Channel Model and Distance Distributions}

The received power at a given user in cell $l$ from its corresponding BS $l$
is defined as follows: 
\begin{equation}\label{gamma}
\gamma_l=P_t r_l^{-\beta} \zeta,
\end{equation}
where $\beta$ is the path-loss exponent, $r_{l}$ is the distance\footnote{Users in a cell are uniformly distributed with respect to their nearest BS  located in the cell-center, i.e., the local users of an active BS $k$ and the sleeping BS $j$ are uniformly distributed with respect to the BS $k$ and $j$, respectively. 
Thus, the distance realizations of all users in a cell $k$ are completely independent of the distance realizations of all users in any other cell $j$, i.e., $r_k$ is completely independent of $r_j$.} of a user in cell $l$ from its corresponding  BS $l$ (see \figref{Multi-Tier} for graphical illustration), $P_t$ is the  transmission power of any arbitrary BS per channel, and $\zeta$ represents the RV  to model composite shadowing and fading channel. Since the users are uniformly distributed in each cell, the distribution of  the distance $r_{l}$ can be  given as follows:
\begin{equation}\label{fr}
f_{r_l}(r)= \frac{2 r}{R^2},
\end{equation}
where $0\leq r \leq R$.  The received power of a  typical user in sleeping cell $j$ from an active BS $k$, which is located at a distance $D_{jk}$ from its previously serving BS, can  be given as:
\begin{equation}
\label{X}
X_{jk}=P_t \, \tilde{r}_{jk}^{-\beta}\, \chi,  \:\:\:\:\:\:\forall k \neq j,
\end{equation}
where $\tilde{r}_{jk}$ is the distance between the user in sleeping cell $j$ and active BS $k$ (see \figref{Multi-Tier} for graphical illustration)\footnote{$\gamma_k$ (a function of  $r_k$) is independent of $X_{jk}$ (a function of $\tilde{r}_{jk}$ which depends on $r_j$ and $D_{jk}$).}, and $\chi$ is a RV to model the corresponding composite shadowing and fading channel. The distribution of the distance $\tilde{r}_{jk}$ can then be given as follows \cite{adelantado2007nonuniform}:
\begin{equation}\label{frtilde}
f_{\tilde{r}_{jk}}(\tilde{r})= \frac{\tilde{r}}{R^2} -\frac{2 \tilde{r}}{\pi R^2}\mathrm{sin}^{-1} \left(\frac{-R^2+D_{jk}^2+\tilde{r}^2}{2 \tilde{r} D_{jk}}\right),
\end{equation}
where $D_{jk}-R \leq \tilde{r} \leq D_{jk}+R$. Note that, the distance $D_{jk}$ depends on the sleeping BS $j$ whose user is under consideration and reference active BS $k$.  Thus, $D_{jk}$ is a variable and it can take any values to incorporate multiple tiers of macrocells. Given a BS switching-off pattern and the sleeping cell $j$ of interest, $D_{jk}$ can be determined from all active BSs and the corresponding distributions of $\tilde{r}_{jk}$  and $X_{jk}$ can be obtained accordingly (see \figref{Multi-Tier}). For instance, if the central BS is sleeping and an active BS is located in second, third, or $n^{\mathrm{th}}$ tier,  $D_{jk}$ will change to $4R, 6R$, or $2 n R$ accordingly.


In general, shadowing and fading channels can be jointly modeled  by composite fading distributions. 
The generalized-$\mathcal{K}$  distribution has been proposed recently in which the shadowing as well as  fading channels are modeled by the  Gamma distribution \cite{bithas,KG}.
As the PDF, CDF, and MGF of the Generalized-$\mathcal{K}$ distribution involve computation-intensive special functions such as \texttt{Meijer-G} and \texttt{Whittaker} functions, we  approximate the distribution with a  more tractable Gamma distribution using the moment matching method, i.e.,
$\mathcal{K}_G(m_c,m_s,\Omega)\approx \mathrm{Gamma}(\kappa,\Theta)$\cite{KG}.
By matching the first and second moments of the two distributions, the corresponding values of $\kappa$ and $\Theta$ can be given as~\cite{KG}:
\begin{equation}\label{approximation}
\kappa=\frac{m_c m_s}{m_c+m_s+1-m_c m_s \epsilon},
\Theta=\frac{\Omega}{\kappa},
\end{equation}
where $\epsilon$ is the adjustment factor.
Thus, $\zeta$ and $\chi$  will be considered as Gamma  RVs throughout the paper. 

\subsection{Methodology of Analysis}
The main steps to characterize the  system performance with BS sleeping, user association, and scheduling are as follows:
\begin{itemize}
\item {\em Step (i)}: Derive the distribution of the received signal power $\gamma_k$ at any arbitrary user in cell $k$ from its corresponding BS $k$.
\item {\em Step (ii)}: Derive the distribution of the received signal power $X_{jk}$ at any arbitrary user in sleeping cell $j$ from an active BS $k$.
\item {\em Step (iii)}:  Derive the best-case mean channel access probability $\tilde{p}_{jk}$ of a user in sleeping cell $j$ with BS $k$ considering both greedy  and round-robin  scheduling schemes, i.e., $\tilde{p}^{\mathrm{GR}}_{jk}$ and $\tilde{p}^{\mathrm{RR}}_{jk}$, respectively.
\item {\em Step (iv)}: Derive the association probability $\hat{p}_{jk}$ of a user in sleeping cell $j$ with BS $k$, considering both MMAP  and MRSP-based user association schemes, i.e., $\hat{p}^{\mathrm{MMAP}}_{jk}$ and $\hat{p}^{\mathrm{MRSP}}_{jk}$, respectively.
\item {\em Step (v)}: Given that  a user in sleeping cell $j$ is associated with BS $k^*$ depending on the user association scheme, derive its exact channel access probability with BS $k^{*}$ (i.e., $p_{jk^*}$), the MGF of its received signal power $S_{jk^*}$  and cumulative interference power $I_{jk^*}^{\mathrm{cum}}$.
\item {\em Step (vi)}: Derive the spectral efficiency  and outage probability of a typical user in sleeping cell $j$.
\end{itemize}

\section{Best-Case Mean Channel Access Probability with Greedy and Round-Robin Scheduling}

In this section, we derive the best-case mean channel access probability of a typical user in a sleeping cell $j$ from an active BS $k$ considering greedy and round-robin scheduling schemes. In this context, first, we derive the exact PDF and CDF of  $\gamma_k$ ({\em Step~(i)}) and then derive the exact PDF, CDF, and MGF of  $X_{jk}$ ({\em Step~(ii)}). We then develop an  approach to approximate the PDF, CDF, and MGF of $X_{jk}$.  Based on this, we finally derive the best-case channel access probability of an arbitrary sleeping cell user  in cell $j$ with respect to an active BS $k$ ({\em Step~(iii)}).

\subsection{Statistical Characterization of $\gamma_k$}
The distribution of the received signal power $\gamma_k$ at any arbitrary user from its serving BS  can be derived  by conditioning on the distribution of $r_k$ in \eqref{gamma}, doing transformation of RV, i.e., $f_{\gamma_k}(\gamma|r_k)= \frac{r_k^\beta}{P_t} f_\zeta\left(\gamma \frac{r_k^\beta}{P_t}\right)$, and averaging over the distribution of $r_k$ as follows:
\begin{equation}\label{fgamma}
f_{\gamma_k}(\gamma)=\int_0^R \frac{r^\beta}{P_t} f_\zeta\left(\gamma \frac{r^\beta}{P_t}\right) f_{r_k}(r) dr.
\end{equation}
Since $\zeta$ is a Gamma RV, i.e., $\zeta \sim \mathrm{Gamma}(\kappa_\zeta,\Theta_\zeta)$, we can re-write \eqref{fgamma} by substituting \eqref{fr} as follows:
\begin{align}\label{fgamma1}
f_{\gamma_k}(\gamma)&=\int_0^R \frac{r^\beta}{P_t} 
e^{-\frac{\gamma r^\beta}{P_t \Theta_\zeta}} \left(\gamma \frac{r^\beta}{P_t}\right)^{\kappa_\zeta-1}
\frac{2 r}{\Gamma(\kappa_\zeta) {\Theta_\zeta}^{\kappa_\zeta} R^2} dr,
\nonumber\\&=
\frac{2 \gamma^{\kappa_\zeta-1}}{\Gamma(\kappa_\zeta) {\Theta_\zeta}^{\kappa_\zeta} R^2 P_t^{\kappa_\zeta}}
\int_0^R r^{\kappa_\zeta \beta+1} 
e^{-\frac{\gamma r^\beta}{P_t \Theta_\zeta}} dr.
\end{align}
After solving the integral in \eqref{fgamma1} using \cite[Eq. 3.381/8]{book}, the closed-form  PDF of $\gamma_k$ can be expressed as
\begin{equation}\label{pdfgamma1}
f_{\gamma_k}(\gamma)=
\frac{2 (P_t \Theta_\zeta)^{2/\beta} }{\beta \Gamma(\kappa_\zeta) \gamma^{1+2/\beta} R^2}
\Gamma_l\left(\kappa_\zeta+\frac{2}{\beta},\frac{R^\beta \gamma}{P_t\Theta_\zeta}\right).
\end{equation}
Consequently, using the property $\Gamma_l(\cdot)+\Gamma_u(\cdot)=\Gamma(\cdot)$ and \cite[Eq.~(06.06.21.0002.01)]{BibWolfram2010Book} and after some algebraic manipulations, the CDF of $\gamma_k$, i.e., $F_{\gamma_k}(\gamma)=\int_0^{\gamma} f_{\gamma_k}(u) du$, can be derived in a closed-form as:
\begin{align}\label{cdfgamma1}
F_{\gamma_k}(\gamma)=
\frac{\Gamma_l(\kappa_\zeta,\frac{R^\beta \gamma}{P_t \Theta_\zeta})}{\Gamma(\kappa_\zeta)}
-\frac{(\Theta_\zeta P_t/\gamma)^{2/\beta} \Gamma_l\left(\kappa_\zeta+\frac{2}{\beta},\frac{R^\beta \gamma}{P_t \Theta_\zeta}\right)}{\Gamma(\kappa_\zeta) R^2}.
\end{align}

\subsection{Statistical Characterization of $X_{jk}$: An Exact Approach}

The PDF of the received signal power  at a sleeping cell user from an active BS $k$, i.e., $X_{jk}$, can be derived similarly by conditioning on the distribution of $\tilde{r}_{jk}$ in \eqref{X}, doing transformation of RV, i.e., $f_{X_{jk}}(x|\tilde{r}_{jk})= \frac{{\tilde{r}_{jk}}^\beta}{P_t} f_\zeta\left(\gamma \frac{{\tilde{r}_{jk}}^\beta}{P_t}\right)$, and finally averaging over the distribution of ${\tilde{r}_{jk}}$ as follows:
\begin{equation}\label{fx1}
f_{X_{jk}}(x)= \int_{A}^{B} \frac{\tilde{r}^\beta}{P_t}
f_\chi \left(x\frac{\tilde{r}_{jk}^\beta}{P_t}\right)  f_{\tilde{r}_{jk}}(\tilde{r}_{jk}) d\tilde{r}_{jk},
\end{equation}
where $A=D_{jk}-R$ and $B=D_{jk}+R$.
Since $\chi$ is a Gamma RV, i.e., $\chi \sim \mathrm{Gamma}(\kappa_\chi, \Theta_\chi)$, by substituting \eqref{frtilde}, we can write \eqref{fx1}  as follows:
\begin{equation}\label{fx2}
f_{X_{jk}}(x)= 
\frac{
\int_{A}^{B} 
\tilde{r}^{\beta\kappa+1} e^{-\frac{x\tilde{r}^\beta}{\Theta P_t}}
\left(1 -
\frac{2}{\pi}\mathrm{sin}^{-1} \left(\frac{D_{jk}^2+\tilde{r}^2-R^2}{2 \tilde{r} D_{jk}}\right)
d\tilde{r}
\right)
}{x^{1-\kappa} P_t^\kappa \Gamma(\kappa) \Theta^\kappa R^2}.
\end{equation}
For notational convenience, we omit the subscripts of $\kappa_\chi$ and $\Theta_\chi$ in this subsection. Using the identity $\int_u^v=\int_0^v-\int_0^u$ and applying \cite[3.381/8]{book} on the first term, \eqref{fx2} can be simplified as follows:
\begin{multline}\label{fx3}
f_{X_{jk}}(x)= 
\frac{ \Gamma\left(\kappa+\frac{2}{\beta},
\frac{A^\beta x}{\Theta P_t},\frac{B^\beta x}{\Theta P_t}\right)
}
{(x/P_t\Theta)^{2/\beta} x\Gamma(\kappa)  R^2 \beta}
\\-
\frac{2 
\int_{A}^{B} 
 {e^{-\frac{x\tilde{r}^\beta}{\Theta P_t}}}{\tilde{r}^{\beta\kappa+1}}
\mathrm{sin}^{-1} \left(\frac{-R^2+D_{jk}^2+\tilde{r}^2}{2 \tilde{r} D_{jk}}\right)
d\tilde{r}
}{x^{1-\kappa} \pi P_t^\kappa \Gamma(\kappa) \Theta^\kappa R^2}.
\end{multline}
Applying Maclaurin series expansion (i.e.,
$
\mathrm{sin}^{-1} y=\sum_{n=0}^\infty  \frac{\Gamma(n+\frac{1}{2})}{\sqrt{\pi} (2 n+1) n!} y^{2 n+1}
$)
and  Binomial expansion, the second term of \eqref{fx3} can be simplified as
$
\sum_{n=0}^\infty
\sum_{m=0}^{2n+1}
\frac{ C  }{\Gamma(\kappa) x {(P_t  \Theta/x)}^\kappa }
\int_{A}^{B} 
\tilde{r}^{\beta\kappa-2 n+2 m} e^{-\frac{x\tilde{r}^\beta}{\Theta P_t}}
d\tilde{r},
$
where
$
C=\frac{2 {2n+1 \choose m} (D_{jk}^2-R^2)^{2n+1-m} \Gamma(n+\frac{1}{2}) }
{\pi \sqrt{\pi} (2 n+1) n! (2 D_{jk})^{2 n+1}  R^2}.
$
Finally, using the identity \cite[Eq. 3.381/8]{book}, the second term of \eqref{fx3} can be solved in closed-form and $f_{X_{jk}}(x)$ can be derived as follows:
\begin{multline}\label{fxfinalpdf}
f_{X_{jk}}(x)=
\frac{ \Gamma\left(\kappa+\frac{2}{\beta},
\frac{A^\beta x}{\Theta P_t},\frac{B^\beta x}{\Theta P_t}\right)}
{(x/P_t\Theta)^{2/\beta} x\Gamma(\kappa)  R^2 \beta}
-\\
\sum_{n=0}^\infty \sum_{m=0}^{2 n+1} 
\frac{C \; \Gamma\left(\kappa+\frac{g}{\beta},
\frac{A^\beta x}{\Theta P_t},\frac{B^\beta x}{\Theta P_t}\right)}{\beta x   \Gamma(\kappa) {(x/P_t \Theta)}^{\frac{g}{\beta}}},
\end{multline}
where $g=1-2n+2m$.
Using the property $\Gamma_l(\cdot)+\Gamma_u(\cdot)=\Gamma(\cdot)$ and \cite[Eq.~(06.06.21.0002.01)]{BibWolfram2010Book} and after some algebraic manipulations, the CDF of $X_{jk}$ can be derived as follows:
\begin{multline}\label{cdfexact}
F_{X_{jk}}(x)=
\frac{
\frac{\Gamma\left(\kappa+\frac{2}{\beta},\frac{B^\beta x}{P_t \Theta},\frac{A^\beta x}{P_t \Theta}\right)}{(x/\Theta)^{2/\beta}}
- \frac{\Gamma_l(\kappa,\frac{A^\beta x}{P_t \Theta})}{1/A^2}
+\frac{\Gamma_l(\kappa,\frac{B^\beta x}{P_t \Theta})}{1/B^2}
}{2 R^2 \Gamma(\kappa)}
-\\
\sum_{n=0}^\infty
\sum_{m=0}^{2 n+1} 
C
\frac{
\frac{\Gamma\left(\kappa+\frac{g}{\beta},\frac{B^\beta x}{P_t \Theta},\frac{A^\beta x}{P_t \Theta}\right)}{(x/\Theta)^{g/\beta}}
- \frac{\Gamma_l(\kappa,\frac{A^\beta x}{P_t \Theta})}{A^{-g}}
+\frac{\Gamma_l(\kappa,\frac{B^\beta x}{P_t \Theta})}{B^{-g}}
}{g \Gamma(\kappa)}.
\end{multline}
Using the definition of MGF, i.e., 
$
\mathcal{M}_{X_{jk}}(t)=\int_0^\infty e^{-t x} f_{X_{jk}}(x) dx,
$
and applying the  identity given in \cite[Eq. 6.455/2]{book}, $\mathcal{M}_{X_{jk}}(t)$ can  be derived  as  in \eqref{mgfexact}.
\begin{figure*}[t]
\begin{multline}\label{mgfexact}
\mathcal{M}_{X_{jk}}(t)=
\frac{\:_2F_1[\kappa,\frac{-2}{\beta},1-\frac{2}{\beta}, 
-\frac{t \Theta}{B^{\beta}} ]}{2 R_m^2/B^2}
-\frac{\:_2F_1[\kappa,\frac{-2}{\beta},1-\frac{2}{\beta},- \frac{t \Theta}{A^{\beta}} ]}{2 R_m^2/A^2}
\\-
\sum_{n=0}^\infty \sum_{m=0}^{2n+1}
C
\left(\frac{\:_2F_1[\kappa,\frac{-g}{\beta}, 1-\frac{g}{\beta}, \frac{-t \Theta}{B^{\beta}}]}{g B^{-g}}-
\frac{\:_2F_1[\kappa,\frac{-g}{\beta}, 1-\frac{g}{\beta},  \frac{-t \Theta}{A^{\beta}}]}{g A^{-g}}
\right).
\end{multline}
\normalsize
\hrule
\end{figure*}
The above exact expression of MGF includes hypergeometric function which is implemented in standard mathematical programming softwares such as \texttt{MATHEMATICA}. However, it might be time consuming to compute hypergeometric functions in other softwares such as \texttt{MATLAB}. As such, in the next subsection, we provide an approximate approach to derive the statistics of $X_{jk}$.

\subsection{Statistical Characterization of $X_{jk}$: An Approximate Approach}

The major complexity factor in the exact derivation  comes from the  distance distribution of $\tilde{r}_{jk}$ provided in \eqref{frtilde}. To avoid this complexity, we discretize the sleeping cell region into $\mathcal{Y}$ circular zones  of  equal width and $\mathcal{W}$ equal angular intervals (other discretization approaches can also be used here as mentioned in \cite{TWChina}). 
Conditioning on the location of a sleeping cell user  from its former serving BS (which is now turned-off) as $(r_y, \theta_w)$, its distance from the active BS $k$ can be calculated by using the cosine law as $\tilde{r}_{w,y}=\sqrt{r_y^2+D_{jk}^2-2 r_y D_{jk} \mathrm{cos}(\theta_{jk}-\theta_w)}$, where $\theta_{jk}$ is the angle between the reference x-axis of BS $j$ and the line connecting BS $j$ and BS $k$ (see Fig.~1 for graphical illustration). For a given location $(r_y, \theta_w)$ of a sleeping cell user, the conditional  MGF of $X_{jk}$ can be given  by using the scaling property of MGF
as $\mathcal{M}_{\chi}(P_t \tilde{r}_{w,y}^{-\beta} t)$. However, this distance $\tilde{r}_{w,y}$ is actually a sample point of the complete sample space  of $\tilde{r}_{jk}$. Thus, the unconditional MGF of $X_{jk}$ can be derived by averaging over the complete sample space of $\tilde{r}_{jk}$ as follows:
\begin{equation}\label{mgfapprox}
\mathcal{M}_{X_{jk}}(t)=
\sum_{y=1}^{\mathcal{Y}} \sum_{w=1}^{\mathcal{W}} 
(1-\Theta_\chi t \tilde{r}_{jk}^{-\beta})^{-\kappa_\chi}
P(\tilde{r}_{jk}=\tilde{r}_{w,y}).
\end{equation}
Since a user is uniformly distributed in cell $j$,  the probability of its existence at each point $(r_y, \theta_w)$ is given by $P(\tilde{r}_{jk}=\tilde{r}_{w,y})= \frac{1}{\mathcal{Y}\mathcal{W}}$.  Along the same lines, the CDF of $X_{jk}$ can be derived as follows:
\begin{equation}\label{cdfapprox}
{F}_{X_{jk}}(x)=
\sum_{y=1}^{\mathcal{Y}} \sum_{w=1}^{\mathcal{W}} 
\frac{\gamma(\kappa, \frac{x \tilde{r}_{w,y}^{\beta}}{P_t \Theta})}{\Gamma (\kappa)}
P(\tilde{r}_{jk}=\tilde{r}_{w,y}).
\end{equation}

\begin{figure}[ht]
\begin{center}
\includegraphics[width=3 in]{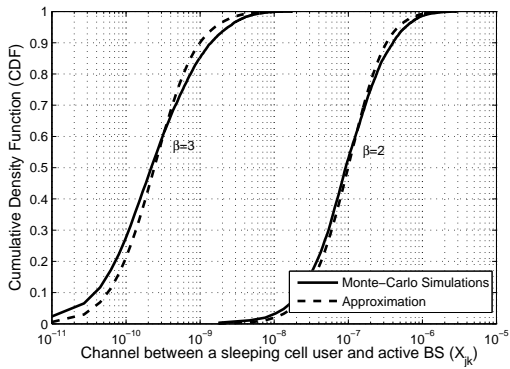}
\caption [c]{
Comparison between the CDF approximation of the received signal power  at a sleeping cell user from an active BS given in Eq. (18)  with the CDF obtained from Monte-Carlo simulations (for $R=500$, $\mathcal{Y}$ = 50, $\mathcal{W}=10$, $\kappa_\chi$ = 2, $\Theta_\chi$ = 4).}
\label{comp-cdf}
\end{center}
\end{figure}

A comparison of the approximated CDF of $X_{jk}$ with that obtained from the Monte-Carlo simulations is demonstrated in \figref{comp-cdf} for different values of $\beta$. It is observed that the discretization approach with a given number of circular zones $\mathcal{Y}$ and angular intervals $\mathcal{W}$ in $r$ and $\theta$, respectively, becomes more accurate for low values of $\beta$. With increasing $\beta$, the signal strength decays much rapidly and the discrete intervals miss large variations of the signal strength. Thus, the divisions especially across $r$, need to be increased for better accuracy, or non-uniform divisions can also be implemented as described in \cite{TWChina}. 

For multiple tiers of macrocells or multiple sleeping BSs, the expression for $X_{jk}$ will change depending on the location of the sleeping cell of interest $j$ and the distances of the active BSs from this BS $j$. The expressions for $X_{jk}$ can be updated by changing the values of $D_{jk}$ accordingly as illustrated in Fig.~1. Note that the derived expressions for $X_{jk}$ consider non-zooming mode of BS $k$; however, all the expressions can be extended to consider zooming mode by  replacing $P_t$ with $\tilde{P}_t=\alpha P_t$.
The expressions for the statistics of $X_{jk}$ will also be useful in deriving the interference power received at a user of sleeping cell $j \in \mathcal{S}$ 
from an active BS $k$. This is due to the fact that the same distributions of the distance and interfering composite fading channel apply  for the received signal powers  from the interfering BSs. 

\subsection{Best-Case Mean Channel Access Probability}

The mean channel access probability is defined as the probability of a  user in sleeping cell $j$ to obtain a channel from active BS $k$ for transmission, when spatially averaged over the area of the sleeping cell. It depends on the user scheduling criterion as well as the number of users in the active cell. Note that, this channel access probability considers only the local users of BS $k$ and is calculated to be used as an association criterion for the users in sleeping cells when MMAP-based association is used. Since more than one sleeping cell users can associate to a BS,  this probability reflects the best-case  and is therefore referred to as the best-case mean channel access probability.

\begin{lemma}[Best-case mean channel access probability for a user in sleeping cell $j$ with greedy scheduling of users at the active BS $k$]
A given active BS allocates a channel to a sleeping cell user $m$ of cell $j$ if its received signal power $X_{jk}^{(m)}$  is greater than the received signal power of its own local users $\gamma_k^{(i)},\:\:\:\forall i=1,2,\cdots,U_k$. 
The  channel access probability corresponding to BS $k$  can then be derived as follows:
\begin{align}\label{accessprob}
\tilde{p}_{jk}^{\mathrm{GR}}&={\mathrm{Pr}}\left({X^{(m)}_{jk}} > 
\underset{\substack{i=1,2,\cdots,U_k}}{\gamma_k^{(i)}}  
\right),
\nonumber\\&=
\int_0^{\infty} \prod_{i=1}^{U_k} {\mathrm{Pr}}(\gamma_{k}^{(i)} \leq X^{(m)}_{jk})
f_{X^{(m)}_{jk}}(x) dx,
\nonumber\\&=
\int_0^{\infty} \prod_{i=1}^{U_k}  F_{\gamma_{k}^{(i)}}(x) f_{X^{(m)}_{jk}}(x) dx.
\end{align}
\normalsize
Considering  independent and identically distributed (i.i.d.) composite fading channel  $\zeta$ of  all users in $k^{\mathrm{th}}$ active cell, the expression for best-case mean channel access probability in \eqref{accessprob} can be simplified as follows\footnote{From this point onward, we will consider only i.i.d. case of $\gamma_k$ for simplicity of expressions. However, this is not a limitation and  the expressions for non-identical case can be obtained in a straight-forward manner.}:
\begin{equation}\label{accessprobgr}
\tilde{p}_{jk}^{\mathrm{GR}}= \int_0^{\infty}  \left( F_{\gamma_{k}}(x)\right)^{U_k} f_{X^{(m)}_{jk}}(x) dx.
\end{equation}
\end{lemma}
\noindent The best-case mean channel access probability in \eqref{accessprob} and \eqref{accessprobgr} can be evaluated using any standard mathematical software packages such as \texttt{MAPLE} and \texttt{MATHEMATICA}. 

\begin{lemma}[Best-case  mean channel access probability of a typical user in a sleeping cell $j$ for round-robin scheduling of users at the active BS $k$]
With round-robin scheduling,  since all users have an equal  chance to access the transmission channel regardless of their channel conditions, $\tilde{p}_{jk}^{\mathrm{RR}}$  can be written as:
\begin{equation}\label{accessprobrr}
\tilde{p}_{jk}^{\mathrm{RR}}=\frac{1}{U_k+1}.
\end{equation}
\end{lemma}
\noindent 
Since $\tilde{p}_{jk}^{\mathrm{RR}}$ reflects only the traffic loads of the active BSs, the users may associate to far away BSs. On the other hand,  $\tilde{p}_{jk}^{\mathrm{GR}}$ is sensitive to both the traffic loads and distances of the active BSs. Thus, for greedy scheduling,  a BS with the highest access probability is not necessarily the BS with the lowest traffic load (for graphical illustration of the MMAP-based association of different sleeping cell users with greedy  and round-robin scheduling schemes, refer to second part of Fig. 1).

\section{Received Signal and Interference Powers at a Sleeping Cell User}

In this section,  we first derive the association probability of a sleeping cell user with different active BSs considering MMAP and MRSP-based user association schemes ({\em Step~(iv)}). For each association scheme, we  show how the exact access probability of a user in sleeping cell $j$ can be calculated for BS $k^*$ with which it is associated. Next, we derive the statistics of the received signal and interference powers at the sleeping cell user in different scenarios ({\em Step~(v)}).

\subsection{Derivations of Association Probability and Exact Channel Access Probability}

\begin{definition}[\textbf{M}aximum Best-Case \textbf{M}ean \textbf{A}ccess \textbf{P}robability (MMAP)-based user association]
A typical  user in a sleeping cell $j$ selects a BS $k^*$ with maximum best-case mean access probability derived in \eqref{accessprobgr} for greedy scheduling   and \eqref{accessprobrr} for round-robin  scheduling.
The user association criterion can be written mathematically as
\begin{equation}
k^{*}=\mathrm{arg\,max}\{\tilde{p}^{(\cdot)}_{jk}\}, \:\:\:k \in \mathcal{A},
\end{equation}
\end{definition}
\noindent
where $(\cdot)$ may be equal to $\mathrm{GR}$ or $\mathrm{RR}$ depending on the scheduling scheme employed at active BSs $k \in \mathcal{A}$. The probability of user association with a given BS $k$ can therefore  be expressed as:
\begin{equation}
\hat{p}_{jk}^{\mathrm{MMAP}}=
\begin{cases}
1, & k=k^{*}\\
0, & \mathrm{else.}
\end{cases}
\end{equation}

That is, a user in a sleeping cell $j$ associates to the BS $k^*$ which provides the highest value of the best-case mean channel access probability. The users in different sleeping cells may associate to different BSs depending on their corresponding mean channel access probabilities with those BSs. The exact channel access probability of a user in sleeping cell $j$ depends on the scheduling scheme at the active BS $k^*$ as well as the number of sleeping cells and their corresponding users associated to BS $k^*$. Since the best-case mean channel access probabilities corresponding to all active BSs can be computed using \eqref{accessprobgr} and \eqref{accessprobrr} for the users in all sleeping cells, their association decisions can be obtained. Consequently, a set of  sleeping BSs  whose users associate to BS $k^*$ can be determined and this set is denoted as $\mathcal{F}_{k^*}$. Note that, $\mathcal{F}_{k^*}$ includes  cell $j$ whose user is under consideration. 

\begin{lemma}[Exact channel access probability of a user in sleeping cell $j$ with MMAP-based user association] 
For greedy scheduling employed at BS $k^*$,  the channels of all users in all sleeping cells that  associate  with BS $k^*$, i.e., $X_{fk^*},\: \:\:\:\forall f \in \mathcal{F}_{k^*}$ need to be compared with  the user of interest $m$ in cell $j$.  The exact channel access probability of a user $m$ in cell $j$ can then be derived for greedy and round-robin scheduling schemes employed at BS $k^*$, respectively, as follows:
\small
\begin{align}\label{exact1}
{p}_{jk^*}^{\mathrm{GR-MMAP}}= &{\mathrm{Pr}}
\left({X^{(m)}_{jk^*}} > 
\left\{
\underset{i=1,2,\cdots,U_{k^*}}{\gamma_{k^*}^{(i)}},
\underset{\substack{n= 1,2,\cdots,U_f, \forall f \in \mathcal{F}_{k^*},\\ 
n\neq m \bigwedge f=j }}{X^{(n)}_{fk^*}}
\right\}
\right),
\\=&
\int_0^{\infty} 
\left( F_{\gamma_{k}}(x)\right)^{U_{k^*}}
\prod_{\substack{n=1,f\in \mathcal{F}_{k^*} \\n\neq m \bigwedge f=j}}^{U_{f}} 
F_{X_{fk^*}^{(n)}}(x)  f_{X^{(m)}_{jk^*}}(x) dx,
\\
\label{rrpkaccess}
\normalsize
p^{\mathrm{RR-MMAP}}_{jk^{*}}=&\frac{1}{U_{k^{*}}+\sum_{f \in \mathcal{F}_{k^*}} U_f},
\end{align}
\end{lemma}
\noindent
where the condition $(n\neq m \bigwedge f=j)$ 
restricts that if $f=j$, then $n \neq m$; this
excludes user $m$ of cell $j$ who is the user of interest.
Note that the product sign shows that $X_{fk^*}$ can be different for different sleeping BSs $f$ depending on their distances $D_{fk^*}$ from the BS $k^*$; thus, the i.i.d. condition may not be applicable for $X_{fk^*}$. Moreover, if all active BSs implement round-robin scheduling, the  channel access probability  depends only on the number of users. In this case, all sleeping cell users may  associate to  a single BS with the lowest number of users. Thus, a user from sleeping cell $j$ will have the following channel access probability from cell $k^{*}$:
\begin{equation}
\label{rrpkaccess}
p^{\mathrm{RR-MMAP}}_{jk^{*}}=\frac{1}{U_{k^{*}}+\sum_{j \in \mathcal{\mathcal{S}}} U_j}.
\end{equation}
\begin{definition}[\textbf{M}aximum \textbf{R}eceived \textbf{S}ignal \textbf{P}ower (MRSP)-based user association]
A typical  user of sleeping cell $j$ selects a BS with maximum received signal power, irrespective of  its corresponding channel access probability.
The criterion of user association can be written mathematically as follows: 
\begin{equation}
k^{*}=\mathrm{arg\,max}\{X_{jk}\},\:\:\:k\in \mathcal{A}.
\end{equation}
\end{definition}
\noindent
With MRSP-based user association, the selected BS $k^{*}$ can be different for a typical user depending on its location within cell $j$. For a typical user of sleeping cell $j$, the association probability with any BS $k$ can be derived as:
\begin{align}\label{ass01}
\hat{p}_{jk}^{\mathrm{MRSP}}&=\mathrm{Pr}(k=k^*)=\mathrm{Pr}(X_{jk}>\underset{\substack{l \in \mathcal{A}\backslash k}}{X_{jl}}),
\nonumber\\&=\int_0^\infty \prod_{\substack{l \in \mathcal{A}\backslash k}} F_{X_{jl}}(x) f_{X_{jk}}(x) dx,
\end{align}
\noindent
where the condition $l \in \mathcal{A} \backslash k$ denotes all active BSs excluding the active BS $k$ and the product sign shows that $X_{jl}$ can vary for different active cells $l$ depending on their distances $D_{jl}$ from cell $j$; thus, the i.i.d. condition may not hold for $X_{jl}$.

Conditioning on that a typical user in sleeping cell $j$ is associated to BS $k^*$, the exact channel access probability depends on the number of users in its own cell as well as the users in other sleeping cells and their corresponding association probabilities with BS $k^*$. 
For the users in all sleeping cells, the association probability with BS $k^*$ depends on their received signal powers from BS $k^*$, i.e., $X_{jk^*}$ and it can be calculated by using \eqref{ass01}. 

The exact channel access probability of the user of interest in cell $j$ can then be derived by
considering a binary vector ${\bf b}_{k^*}$ of cardinality $\sum_{j \in \mathcal{S}} U_j-1$ in which each bit represents the state of a sleeping cell user, i.e., $b_{k^*}(i)=1$ if a user is associated to BS $k^*$ and $b_{k^*}(i)=0$ otherwise. Each binary vector  represents the state of all users in all sleeping cells excluding the user of interest $m$ in cell $j$. The set of all possible combinations of ${\bf b}_{k^*}$ is denoted as $\mathcal{B}$. 
The probability of each possible  combination ${\bf b}_{k^*}$  can be calculated as
\begin{equation}\label{comb}
\mathrm{Pr}({\bf b}_{k^*})=\prod_{i=1}^{\sum_{j \in \mathcal{S}} U_j-1} \left(\hat{p}_{lk^*}^{\mathrm{MRSP}} \right)^{b_{k^*}(i)}(1-\hat{p}_{lk^*}^{\mathrm{MRSP}})^{1-b_{k^*}(i)},
\end{equation} 
where $\hat{p}_{lk^*}^{\mathrm{MRSP}}$ is the association probability of any  user $n$ in sleeping cell $l \in \mathcal{S}$ with BS $k^*$ such that $n=1,2,\cdots,U_l$.  The condition $l=j \bigwedge n\neq m$ excludes user $m$ of sleeping cell $j$.
Given the probability of each combination as derived in \eqref{comb}, the exact access probability  for the typical user in cell $j$ can then be derived as follows.
\begin{lemma}[Exact access probability of a user in a sleeping cell $j$ with MRSP-based user association] 
For a given combination ${\bf b}_{k^*}$, the number of associated users with BS $k^*$ can be given as $\sum_{i=1}^{\sum_{j \in \mathcal{S}} U_j-1} b_{k^*}(i)$. For greedy scheduling, the channels of the corresponding associated users in different sleeping cells need also be considered.
The exact channel access probability can be derived for  greedy and round-robin scheduling schemes at BS $k^*$,  respectively, as follows:
\begin{align}\label{exact2}
p_{jk^*,\bf{b}_{k^*}}^{\mathrm{GR-MRSP}}&
{=}
\int_0^{\infty}  
\left( F_{\gamma_{k^*}}(x)\right)^{U_{k^*}}
\times\nonumber\\&
\prod_{\substack{i=1, l \in \mathcal{S} \\ l=j \bigwedge n\neq m}}^{\sum_{j \in \mathcal{S}} U_j-1}
\left(F_{X^{(n)}_{lk^*}}(x)\right)^{b_{k^*}(i)}
f_{X^{(m)}_{jk^*}}(x) dx, 
\\
\label{exactrr}
p^{\mathrm{RR-MRSP}}_{jk^*,\bf{b}_{k^*}}&=
\frac{1}{U_{k^*}+\sum_{i=1}^{\sum_{j \in \mathcal{S}} U_j-1} b_{k^*}(i)+1},
\end{align}
where $n=1,2,\cdots,U_l$.
The  unconditional channel access probability can then be derived by considering all possible combinations of  set $\mathcal{B}$ and summing them up, i.e., $p_{jk^*}^{\mathrm{GR-MRSP}}=\sum_{\bf{b}_{k^*} \in \mathcal{B}} p_{jk^*,\bf{b}_{k^*}}^{\mathrm{GR-MRSP}} \mathrm{Pr}(\bf{b}_{k^*})$ and $p_{jk^*}^{\mathrm{RR-MRSP}}=\sum_{\bf{b}_{k^*} \in \mathcal{B}} p_{jk^*,\bf{b}_{k^*}}^{\mathrm{RR-MRSP}} \mathrm{Pr}(\bf{b}_{k^*})$.
\end{lemma}
\noindent
Based on this, in the next subsection, we will define the received signal and  interference powers at a sleeping cell user and derive their corresponding CDF and MGF, respectively.

\subsection{Statistics of the Received Signal Power at a Sleeping Cell User}

\subsubsection{MMAP-based user association and greedy scheduling}
The received signal power of a typical user in sleeping cell $j$ associated with BS $k^{*}$ can be defined  as:
\begin{equation}
S_{jk^{*}}= 
\begin{cases}
{\mathrm{arg\,max}} \{ \underset{\substack{i=1,2,\cdots, U_{k^*}}}{\gamma_{k^*}^{(i)}}  
\underset{\substack{n = 1,2,\cdots,U_f\\f\in \mathcal{F}_{k^*}}}{X^{(n)}_{fk^*}}\}, & \mathrm{if \:\:\:\:selected}\\
0, & \mathrm{otherwise.}
\end{cases}
\end{equation}
\normalsize
Note that the user will remain silent and have a received signal power of zero if not selected for transmission. The  CDF of $S_{jk^{*}}$ can then be derived as follows:
\begin{align}
F_{S_{jk^{*}}}(s)&=
p^{\mathrm{GR-MMAP}}_{jk^*} (F_{\gamma_{k^*}}(s))^{U_{k^*}} \prod_{\substack{n=1\\f\in \mathcal{F}_{k^*}
}}^{U_f}  F_{X^{(n)}_{fk^*}}(s) 
+\nonumber\\&
(1-p^{\mathrm{GR-MMAP}}_{jk^*})\mathbb{U}(s).
\end{align}
Note that the product sign highlights the fact that the received signal powers of different users  of different sleeping cells are not i.i.d. rather they are independent due to different values of $D_{fk^*}$. 
The MGF of a random variable can be derived from its PDF as well as CDF. However, in this section, we  derive the MGF of $S_{jk^{*}}$ from CDF instead of PDF as follows:
\begin{align}\label{mgfdef}
\mathcal{M}_{S_{jk^{*}}}(t)=& 
\int_0^\infty t e^{-t S} F_{S_{jk^*}}(S) dS,
\nonumber\\
=&
t p^{\mathrm{GR-MMAP}}_{jk^*} 
\int_0^\infty e^{-t S} F_{S_{jk^{*}}}(S)
dS
+\nonumber\\&
(1-p^{\mathrm{GR-MMAP}}_{jk^*}).
\end{align}

\subsubsection{MMAP-based user association and round-robin scheduling}
For round-robin scheduling, the signal power received at a typical user in a sleeping cell who is associated with BS $k^{*}$ is defined  as follows:
\begin{equation}
S_{jk^{*}}= 
\begin{cases}
{X_{jk^*}}, & \mathrm{if \:\:\:\:selected}\\
0, & \mathrm{otherwise.}
\end{cases}
\end{equation}
The CDF of $S_{jk^{*}}$ can then be derived as follows: 
\begin{equation}
F_{S_{jk^{*}}}(s)=  p^{\mathrm{RR-MMAP}}_{jk^*} F_{X_{jk^*}}(s) +(1-p^{\mathrm{RR-MMAP}}_{jk^*})\mathbb{U}(s).
\end{equation}
Thus, the MGF can then be derived  as 
$
\mathcal{M}_{S_{jk^{*}}}(t)= 
p^{\mathrm{RR-MMAP}}_{jk^*}\mathcal{M}_{X_{jk^*}}(t) +(1-p^{\mathrm{RR-MMAP}}_{jk^*}).
$
Note the same MGF definition as in \eqref{mgfdef} can be applied for all cases. Therefore, for conciseness, we will skip the   expressions for MGF and detail only the expressions for CDF in the following.

\subsubsection{MRSP-based user association and greedy scheduling}
Given that a typical user of sleeping cell $j$ associated with BS $k^*$ gets access to the channel, the CDF of the  received  signal power can be derived for a given combination $\bf{b}_{k^*}$ as:
\begin{multline}
F_{S_{jk^{*},\bf{b}_{k^*}}}(s)= 
\left( F_{\gamma_{k^*}}(x)\right)^{U_{k^*}}
\prod_{k=1}^{N_{\mathrm{active}}} F_{X^{(m)}_{jk}}(s) 
\times\\
\prod_{i=1}
^{\sum_{j \in \mathcal{S}} U_j-1}
\left(F_{X^{(n)}_{lk^*}}(x)\right)^{b_{k^*}(i)}.
\end{multline}
\normalsize
Note that the product sign $\prod_{k=1}^{N_{\mathrm{active}}}$  represents that the received signal powers of a user in cell $j$ from all active BSs are also not i.i.d. since $D_{jk}$ can be different for different active BSs. Moreover, $\prod_{i=1}^{\sum_{j \in \mathcal{S}} U_j-1}$ represents that the received signal powers of different sleeping cell users who are associated to BS $k^*$ are also not i.i.d. due to the different values of $D_{lk^*}$.
The unconditional CDF can then be given as follows:  
\begin{multline}
F_{S_{jk^{*}}}(s)=
\sum_{ {\bf b}_{k^*} \in \mathcal{B}} 
\left(
F_{S_{jk^{*},\bf{b}_{k^*}}}(s)
{p}_{lk^*,\bf{b}_{k^*}}^{\mathrm{GR-MRSP}}\right.\\
\left.
+
(1-{p}_{lk^*,\bf{b}_{k^*}}^{\mathrm{GR-MRSP}}) \mathbb{U}(s)
\right) \mathrm{Pr}(\bf{b}_{k^*}).
\end{multline}
  
\subsubsection{MRSP-based user association and round-robin scheduling}
For round-robin scheduling, the signal power received at a typical user in a sleeping cell who is associated with BS $k^*$ can be defined as:
\begin{equation}
S_{jk^{*}}= 
\begin{cases}
{\mathrm{arg\,max}} \{X_{jk}
\}, & \mathrm{if \:\:\:\:selected}, k \in \mathcal{A},\\
0, & \mathrm{otherwise.}
\end{cases}
\end{equation}
The CDF  can then be derived as follows:
\begin{align}\label{mrsp-rr}
F_{S_{jk^{*}}}(s)=&
p^{\mathrm{RR-MRSP}}_{jk^*}  \prod_{k \in \mathcal{A}} F_{X_{jk}}(s)  +(1-p^{\mathrm{RR-MRSP}}_{jk^*} )\mathbb{U}(s). 
\end{align}
Note that, all derived expressions for the channel access probabilities and MGF of the received signal power consider non-zooming mode of BS $k$. However, the impact of zooming can be incorporated by replacing $P_t$ with $\alpha P_t$ in the statistics of $X_{jk}$.

\subsection{Statistics of the Interference Power at a Sleeping Cell User}
In this subsection, we derive the MGF of the interference received at a sleeping cell user for both MMAP and MRSP-based user association schemes. The cumulative interference power received at a typical  user in sleeping cell $j$ associated to $k^*$ depends on the transmit power of the other BSs and their corresponding zooming probabilities. The cumulative interference can then be defined as follows: 
\begin{align}\label{cummgf}
I_{jk^*}^{\mathrm{cum}}= \sum_{\substack{k=1}}^{N_{\mathrm{active}}-\tilde{N}_{\mathrm{zoom}}} {P}_t \tilde{r}_{jk}^{-\beta} \chi+
\sum_{\substack{k=N_{\mathrm{active}}-\tilde{N}_{\mathrm{zoom}}+1}}^{N_{\mathrm{active}}} \alpha {P}_t \tilde{r}_{jk}^{-\beta} \chi,
\end{align}
where $k\in \mathcal{A}, k\neq k^*$ and $\tilde{N}_{\mathrm{zoom}}={N}_{\mathrm{zoom}}-1$ denotes the number of interfering BSs which are increasing their transmit powers. 
Consider a binary vector ${\bf z}$ of size $N_{\mathrm{active}}-1$ in which each bit represents the state of an interfering BS, i.e., $z(k)=1$ if a BS is zooming and $z(k)=0$ otherwise. The probability of  each possible zooming combination  can then  be derived as:
\begin{equation}
\mathrm{Pr}({\bf z})=\prod_{k=1}^{N_{\mathrm{active}}-1}(p_{k}^{\mathrm{zoom}})^{z(k)}(1-p_{k}^{\mathrm{zoom}})^{1-z(k)}.
\end{equation}

For MRSP-based user association, the zooming probability of a BS $k\neq k^*$ can be given as the sum of all probabilities in which any user of sleeping cell $l$ gets access to a transmission channel from BS $k$, i.e., $p_{k}^{\mathrm{zoom}}=\sum_{l \in \mathcal{S}} p_{lk}^{\mathrm{(\cdot)-MRSP}}$ excluding the user of interest in sleeping cell $j$ as this user is associated and selected in BS $k^*$. Similarly, for MMAP-based user association, the zooming probability of BS $k$ can be given as $p_{k}^{\mathrm{zoom}}=\sum_{f \in \mathcal{F}_k} p_{fk}^{\mathrm{(\cdot)-MMAP}}$, where $(\cdot)$ represents $\mathrm{GR}$ or $\mathrm{RR}$ depending on the scheduling scheme used at BS $k$. 

Finally, the MGF of $I_{jk^*}^{\mathrm{cum}}$ for each possible combination can  be derived by applying the scaling law of MGF as:
\begin{equation}
\mathcal{M}_{I_{jk^*,{\bf z}}^{\mathrm{cum}}}(t)
=\prod_{\substack{k=1\\k\neq k^*}}^{N_{\mathrm{active}}-1}
(\mathcal{M}_{X_{jk}}(t))^{1-z(k)}
(\mathcal{M}_{X_{jk}}(\alpha t))^{z(k)}
.
\end{equation}
Finally, by averaging over all possible combinations, the MGF of $\mathcal{M}_{I_{jk^*}}^{\mathrm{cum}}(t)$ can be derived as follows:
\begin{equation}
\mathcal{M}_{I_{jk^*}^{\mathrm{cum}}}(t)=
\sum_{z \in \mathcal{Z}}\mathcal{M}_{I_{jk^*,{\bf z}}^{\mathrm{cum}}}(t) \mathrm{Pr}({\bf z}).
\end{equation} 

\section{Performance Metrics and Design Guidelines for Hybrid User-Association Schemes}

In this section, we demonstrate the significance of the derived  MGF expressions of the cumulative interference  and received signal power at a typical user in sleeping cell $j$ to quantify important network performance metrics such as the outage probability and average spectral efficiency. Moreover, preliminary design guidelines are also provided to develop  hybrid user association schemes that can combine different features of MMAP and MRSP  schemes and assist in improving the system performance even further.

\subsubsection{Spectral efficiency}
Since the two RVs $S_{jk^*}$ and $I_{jk^*}^{\mathrm{cum}}$ are independent, for different scheduling and user association schemes, the spectral efficiency of transmission to a sleeping cell user can be  calculated by using  the lemma proposed in \cite{hamdi} as:
\begin{align}\label{cap}
\mathcal{C}_{jk^*}=&\mathbb{E}\left[\mathrm{ln} \left(1+\frac{S_{jk^*}}{I_{jk^*}^{\mathrm{cum}}+\sigma^2}\right)\right],
\\&=\int_0^\infty \frac{\mathcal{M}_{I_{jk^*}^{\mathrm{cum}}}(t)(1-\mathcal{M}_{S_{jk^*}}(t))
}{t} e^{-\sigma^2 t} dt,
\end{align}
where $\sigma^2$ denotes the thermal noise power. Note that  the  expressions for $\mathcal{M}_{I_{jk^*}^{\mathrm{cum}}}(t)$ and $\mathcal{M}_{S_{jk^*}}(t)$ have been derived in Section~IV. 
The average spectral efficiency  of transmission to a   sleeping cell user can then be calculated as:
\begin{equation}\label{cfinal}
\mathcal{C}_{j}=\sum_{k \in \mathcal{A}} \mathcal{C}_{jk^*} \hat{p}^{(\cdot)}_{jk}.
\end{equation}

\subsubsection{Outage probability}
Similarly, the outage probability, which is defined as the probability of the instantaneous interference-to-signal-ratio to exceed a certain threshold $\mathcal{Q}$, can also be derived by using the characteristic function approach proposed in \cite{qt}.
Given the characteristic function of interference $I_{jk^*}^{\mathrm{cum}}$ and signal power $S_{jk^*}$ and the fact that they are independent, the outage probability is given as
\begin{equation}
\label{QTZhang}
P_{\mathrm{out}}=\frac{1}{2} +\frac{1}{\pi }\int_{0}^\infty \mathrm{Im}\left(\frac{\phi_{I_{jk^*}^{\mathrm{cum}}}(j \mathcal{Q}\omega)\phi_{S_{jk^*}}(-j\omega)}{\omega}\right) d\omega,
\end{equation}
where $\mathrm{Im}(\cdot)$ denotes the imaginary part and $\phi(\cdot)$ represents the characteristic function that can be derived from the MGF. 
Using \eqref{QTZhang}, the outage probability can be evaluated by using any standard mathematical software package such as \texttt{MATHEMATICA}. 

\subsubsection{Extensions to hybrid user association schemes}
A simple hybrid user association criterion  is one in which a sleeping cell user selects a BS dynamically considering its received signal power from all active BSs as well as their corresponding channel access probabilities, e.g., $k^{*}=\mathrm{arg\,max}\{\tilde{p}^{(\cdot)}_{jk} X_{jk}\}$.
This criterion restricts the users from associating to a congested BS despite a high signal power received from that BS.  In this case, the channel access probabilities serve as a weight to the corresponding received signal powers from active BSs and thus  implicitly tend to balance the traffic load among the active BSs. 

Another possible strategy is to select a set of BSs within a certain region of vicinity  and obtain the corresponding channel access probabilities. This can be referred to as location-aware MMAP-based association in which both the locations (and hence the channel  conditions) as well as the access probability will be considered. In this scheme,  there is no need to request the access probabilities of  BSs that are located far away from the sleeping cell users. This scheme is expected to perform significantly better than MMAP-based association with round-robin scheduling. 

\subsubsection{Spectral efficiency of users in active cells}
In this paper, we focus on the spectral efficiency of sleeping cell users. However, 
a similar analysis can also be performed for local users of an active BS $k$. Note that the local users in a BS $k$ remain associated to BS $k$; therefore, we need to derive only the exact access probability of such users after the association of sleeping cell users. If MMAP-based association is used by the sleeping cell users, the number of sleeping cells and their users associated to the active  BS $k$ (i.e., $\mathcal{F}_k$) can be determined. Given the set $\mathcal{F}_k$, the access probability of a local user of BS $k$ can be determined for  greedy scheduling by using a method similar to that described through Eq.~(24) and Eq.~(25) as follows:
\begin{align}\label{exact1}
{p}_{k}^{\mathrm{GR-MMAP}}= &{\mathrm{Pr}}
\left({\gamma^{(m)}_{k}} > 
\left\{
\underset{\substack{i=1,2,\cdots,U_{k}\\ i \neq m}}{\gamma_{k}^{(i)}},
\underset{\substack{n= 1,2,\cdots,U_f\\ \forall f \in \mathcal{F}_{k}
 }}{X^{(n)}_{fk}}
\right\}
\right).
\end{align}
\normalsize

Similarly, if sleeping cell users associate using MRSP-based association, a set of users $\mathcal{B}$  of cardinality $\sum_j U_j$ can be made. For all possible combinations of this set, the access probability of a local user in cell $k$ can be calculated by using a method similar to that  described through Eq.~(31). 
For round-robin scheduling, since each user has equal access probability, the access probability of a local user remains the same as given in Eq.~(26), Eq.~(27), and Eq.(32).
Once the access probabilities are derived, the received signal power by active cell users can also be derived by averaging over the newly derived exact access probabilities for greedy scheduling. On the other hand,  for round-robin scheduling and MMAP-based association of sleeping cell users,  $X_{jk^*}$ needs also be replaced with $\gamma_{k}$ in Eq. (36). With MRSP-based association of sleeping cell users and round-robin scheduling, $\mathrm{argmax} \{X_{jk}\}$ needs to be replaced with $\gamma_{k}$ in Eq.~(40). Similarly,  the interference from all neighboring BSs can be obtained  by considering the received signal powers at a user in cell $k$ from  all active interfering BSs $X_{lk},\:\:\:l\neq k$.

\section{Numerical  Results and Discussions}

In this section,  we  quantify and analyze the  spectral efficiency  of a typical sleeping cell user as  a function of different user association and scheduling schemes. A wide range of  performance trends are captured  to extract insights related to the performance of a typical user in a cell with no BS sleeping, BS sleeping with no cell-zooming, and BS-sleeping with cell-zooming. Moreover, the overall network spectral and energy efficiencies are also captured by defining the  energy efficiency metric as the {\em ratio of  network spectral efficiency [bits/sec/Hz] and network power consumption at the BSs [J/sec]}. The metric is computed by using Monte-Carlo simulations.

\begin{figure}[h]
\begin{center}
\includegraphics[width=3.5 in]{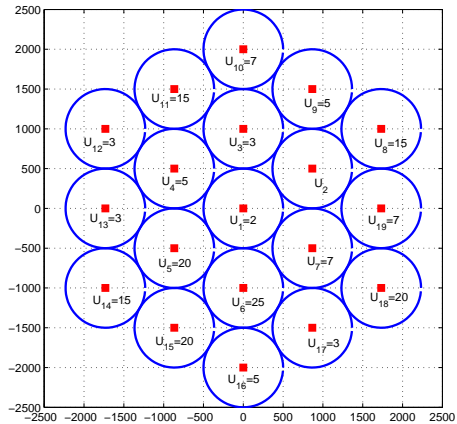}
\caption [c]{Two tiers of macrocell BSs and their corresponding users.}
\label{twotier}
\end{center}
\end{figure}

We consider two tier of macrocells, i.e., $L=19$ with $R_m= 500$~m. The  path-loss exponent $\beta= 2.6$, thermal noise power $\sigma^2=1 \times 10^{-16}$  W/Hz, and initial transmission power per channel $P_t=1$ W. The Monte-Carlo simulation results are averaged over 100,000 iterations. The  static power consumption per BS $P_{\mathrm{static}}=200$~W, $P_{\mathrm{sleep}}=2~W$, and  $\Delta_{\mathrm{dyn}}=3.77$ \cite{arnold2010power}\footnote{In deep sleep modes, a sizable fraction of the hardware in a BS is switched off \cite{powersleepmode1}. Thus, the power consumption in sleep mode  is significantly small compared to the overall BS power consumption (i.e., around 0.5~W \cite[Section~III]{powersleepmode1} or 1~W \cite[Table~I]{powersleepmode2}).  
}. 
The interference and desired channel composite fading are taken as  $f_\chi(\chi)\sim \mathrm{Gamma}(2,1)$ and   $f_\zeta(\zeta) \sim \mathrm{Gamma}(1/2,1)$. In both simulation and analysis, circular macrocells are considered and generalized-$\mathcal{K}$ composite shadowing and fading is approximated by tractable Gamma distribution\footnote{Since the considered approximations are well-known due to their analytical tractability and their accuracies are well-investigated in the prior works \cite{KG,baltzis2011hexagonal},  we only focus on validating the accuracy of the analytical expressions derived in this manuscript. 
}. 
We consider non-uniform traffic load scenario as illustrated in  \figref{twotier}. 

\subsection{Best-Case Mean Channel Access Probability of a Sleeping Cell User}
\begin{figure}[h]
\begin{center}
\includegraphics[width=3.5 in]{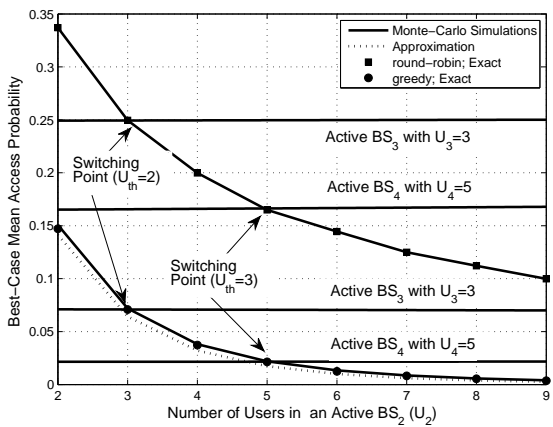}
\caption{Best-case mean channel access probability of a sleeping cell user as a function of the number of users in a given active cell for greedy and round-robin scheduling schemes under non-uniform traffic load scenario.}
\label{1}
\end{center}
\end{figure}

\figref{1} illustrates the best-case mean channel access probability of a user in the sleeping cell (BS$_{1}$)  as a function of the number of users in BS$_2$ with greedy and round-robin scheduling schemes (as derived in \eqref{accessprobgr} and \eqref{accessprobrr}), respectively.   The derived exact expressions match well with those obtained from  Monte-Carlo simulations and the impact of the inaccuracy due to the approximated approach is also observed to be minimal. Compared to the round-robin scheduling, the access probability of a user with active BS$_2$ is significantly low for  greedy scheduling scheme. The main reason is the channel-aware selection of the users in greedy scheduling and the fact that the sleeping cell users  are located beyond the cell-edge of the active BSs.  This results in significantly high chances of selecting a user from within the active cell compared to a sleeping cell user who is located beyond the cell-edge. 
On the other hand, with round-robin scheduling, all  users have an equal chance of accessing the transmission channel. Moreover, since the access probability depends on the traffic load in active BSs, it  decreases as the number of users in  BS$_2$  increases. 
Note that with $U_{\mathrm{th}}=3$, the switching point  for a typical user in the central sleeping cell occurs when $U_2 \geq 5$ since there is another active BS$_4$ with $U_4=5$. Similarly,  with $U_{\mathrm{th}}=2$, the switching  to BS$_3$ with $U_3=3$ occurs immediately when $U_2$ exceeds three.
This switching will help reducing the  degradation of spectral efficiency any further due to an increase in $U_2$ with MMAP-based association.

\subsection{Spectral Efficiency with BS Sleeping and No Cell-Zooming}

\begin{figure}[h]
\centering
\includegraphics[width=3.5in]{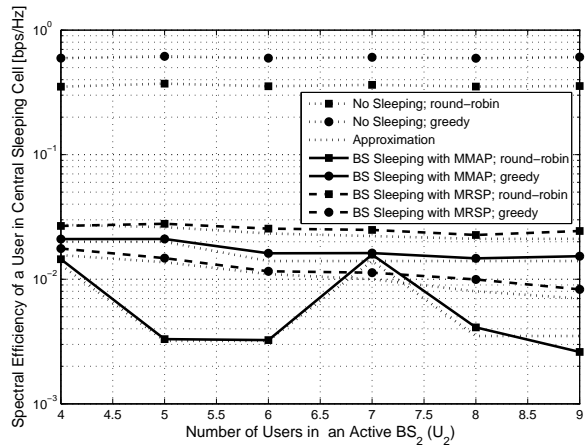}
\caption [c]{Spectral efficiency of transmission to a sleeping cell user before and after BS sleeping with no cell-zooming as a function of the number of users in a given active cell for different user association and scheduling schemes (for $U_{\mathrm{th}}=3$).}
\label{2}
\end{figure}

\begin{figure}[h]
\centering
\includegraphics[width=3.5in]{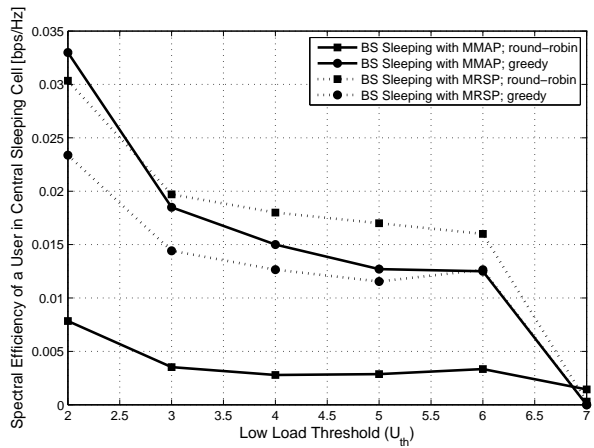}
\caption [c]{Spectral efficiency of transmission to a sleeping cell user  with no cell-zooming as a function of the low load threshold for different user association and scheduling schemes ($U_2=5$).}
\label{2a}
\end{figure}

With greedy and round-robin scheduling schemes at active BSs and $U_{\mathrm{th}}=3$, \figref{2} quantifies the spectral efficiency degradation for a typical user of the central sleeping cell. The results from Monte-Carlo simulations match nearly perfectly with the derived approximation of the spectral efficiency for different user association and scheduling schemes. 
For round-robin scheduling, with MMAP-based association, the spectral efficiency turns out to be significantly low compared to that with MRSP-based association.  This is due to the fact that for round-robin scheduling, the MMAP-based association is independent of channel conditions and a sleeping cell user always associates with the low-loaded BS, regardless of the distance of the corresponding BS. Therefore, MRSP or hybrid/location aware MMAP-based association  schemes, as discussed in Section~V.C, are recommended with round-robin scheduling schemes. Also, note that the spectral efficiency of transmission to a sleeping cell user  switches between the two levels. This is due to the fact that it can associate to a BS with $U=5$ in both first and second tier of macrocells.

For greedy scheduling, the performance of MMAP-based association is better than MRSP-based association. Note that the channel access probability for greedy scheduling is sensitive to both traffic load as well as channel condition; therefore, a typical sleeping cell user will be highly unlikely to associate with far away BSs.
With the increase in $U_2$, the spectral efficiency of transmission to a sleeping cell user degrades; however, MMAP-based association allows these users to switch to another BS with a fixed load of five users (as shown in Fig.~\ref{1}). 


Similar  trends  in spectral efficiency of transmission to the user of interest can also be observed from \figref{2a} as a function of different load thresholds ($U_{\mathrm{th}}$). At $U_{\mathrm{th}}=7$, all cells in the first tier are either turned off or are highly  loaded. Therefore, a sharp performance degradation can be observed for both association schemes.
Given a non-uniform traffic load scenario and a desired spectral efficiency of transmission to a typical sleeping cell user, a suitable value of $U_{\mathrm{th}}$ can be selected.

\begin{figure}
\begin{center}
\includegraphics[width=3.5 in]{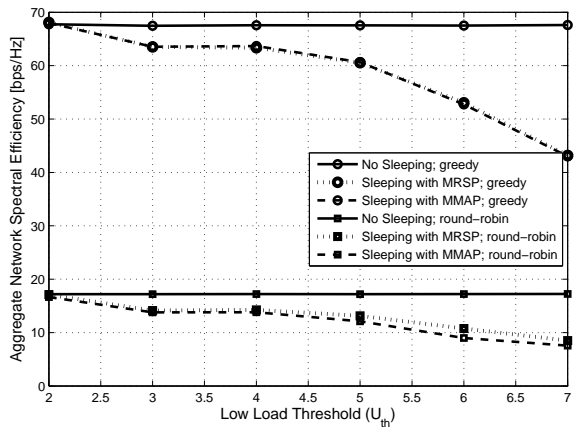}
\caption [c]{Overall system spectral efficiency with no zooming as a function of 
$U_{\mathrm{th}}$ for different user association and scheduling schemes.}
\label{3}
\end{center}
\end{figure}

\figref{3} quantifies the overall network spectral efficiency as a function of  $U_{\mathrm{th}}$ under different user association and scheduling schemes. Compared to the round-robin scheduling, the greedy scheme offers a significantly higher spectral efficiency since it can exploit multi-user diversity(i.e.,  the  number of available users and their corresponding channel gains).
For round-robin scheduling, with an increase in the number of sleeping BSs, the overall network spectral efficiency continues to degrade.

The impact of different  user association schemes of sleeping cell users on the network spectral efficiency is not significant. This is due to low number of the sleeping cell users and  their minimal channel access probability especially with greedy scheduling scheme. 
The difference is slightly more visible for round-robin scheduling at higher values of $U_{\mathrm{th}}$ since the number of sleeping cell users increases  who have relatively higher channel access probabilities compared to greedy scheduling scheme. 
With MMAP-based association and round-robin scheduling, the network spectral efficiency slightly degrades due to the higher chances of selection of sleeping cell users compared to MRSP-based association and their poor channel conditions.

\subsection{Spectral Efficiency with  BS Sleeping and Cell-Zooming}


\begin{figure}[h]
\begin{center}
\includegraphics[width=3.5in]{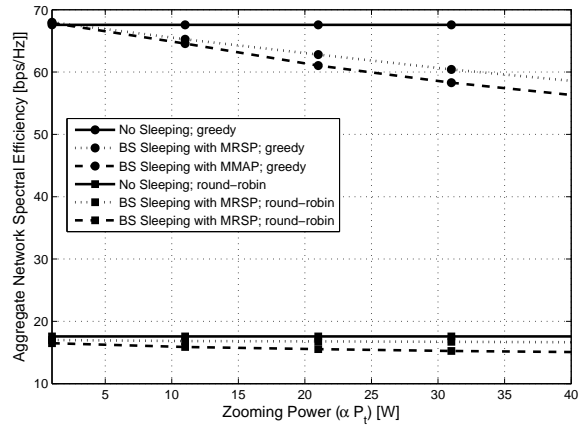}
\caption [c]{Overall system spectral efficiency with zooming as a function of transmit power allocated for zooming purpose for different scheduling and user association schemes (for $U_{\mathrm{th}}=2$, $U_2=5$).}
\label{5}
\end{center}
\end{figure}

\begin{figure}[h]
\begin{center}
\includegraphics[width=3.5 in]{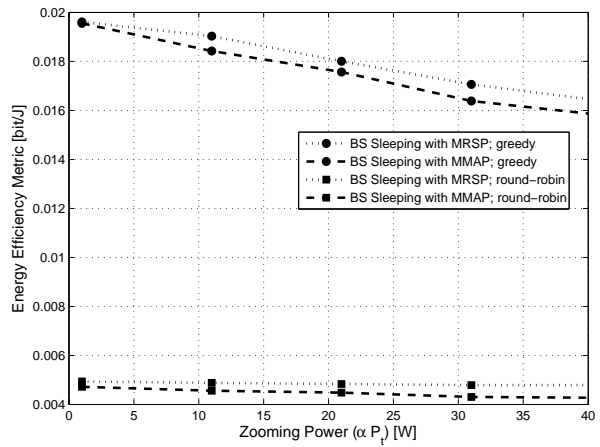}
\caption [c]{Energy efficiency metric as a function of the  transmit power allocated for zooming purpose for different user association and scheduling schemes (for $U_{\mathrm{th}}=2$, $U_2=5$).}
\label{6}
\end{center}
\end{figure}

\figref{5} depicts the overall system spectral efficiency as a function of the increase in available transmit power for users in sleeping cells.  Due to increased interference, the system spectral efficiency is observed to decay with the increase in the zooming power.  
The overall system spectral efficiency is relatively better for MRSP-based association when compared to MMAP-based association. The reason behind this trend is that, with MRSP-based user association, the access probability of sleeping cell users from majority of BSs can be quite low compared to MMAP-based association  in which case a user always  associates to the BS offering the highest channel access probability. This reduces the chance of a BS to zoom and inturn reduces the interference in MRSP-based user association.  While it is advantageous from a system point of view, this has a direct impact on the spectral efficiency of sleeping cell users  which we will see next.  Similar trends for energy efficiency metric,  which is the ratio of network capacity and the total network power consumption, can also be observed in \figref{6}.

\begin{figure}[h]
\begin{center}
\includegraphics[width=3.5 in]{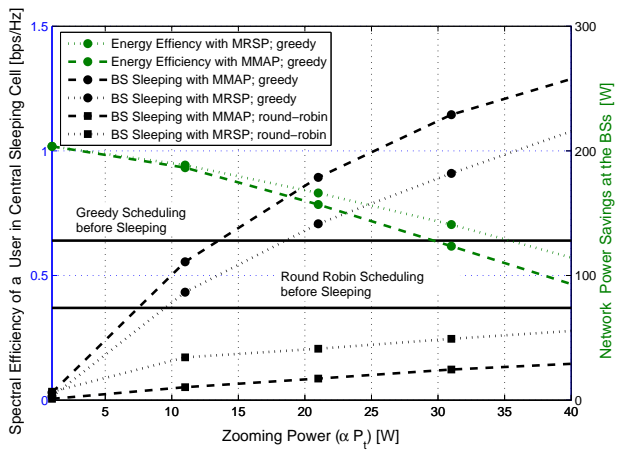}
\caption [c]{Increase in the spectral efficiency of transmission to a sleeping cell user with the increase in transmit power of the selected BS for different scheduling and user association schemes (for $U_{\mathrm{th}}=2$, $U_2=5$).}
\label{4}
\end{center}
\end{figure}

\figref{4} depicts the impact of increasing the transmit power on the  attained spectral efficiency of the sleeping cell users for different scheduling and user association schemes. The spectral efficiency gains of the sleeping cell users are observed to be higher in MMAP-based association with greedy scheduling scheme when compared to the three other possibilities of user association and scheduling.  This is due to the fact that, with greedy scheduling,
the increase in transmit power for zooming  works like a bias for sleeping cell users (i.e., the channel access probability increases).  
However, a trade-off can be observed in the spectral and energy efficiency performances. With MMAP, for a given zooming power,   the  zooming rate increases for the sleeping cell users. This reduces the energy efficiency of MMAP-based association when compared to MRSP-based associatio for a given zooming power. However, it can be seen that the MMAP-based association achieves the target rates with less zooming power of the BSs.
On the other hand, with round-robin scheduling, the access probability of a sleeping cell user  is completely independent of the increase in transmit power. Therefore,  round-robin scheduling  tracks the target much slowly and may not be able to achieve its target with a reasonable energy efficiency.

\section{Conclusion}

A framework has been developed to analyze the performance of BS sleeping while capturing the impacts of different  scheduling  and  user association schemes.  In non-uniform traffic load scenarios, MMAP-based user association enhances the performance of sleeping cell users with greedy scheduling of users at the active BSs. On the other hand, MRSP-based user association improves their performance with round-robin scheduling. User association and scheduling impact the zooming power required by the active BSs to provide coverage to sleeping cell users. To achieve a good trade-off between spectral and energy efficiencies,  a BS should zoom as low as possible  to avoid the degradation of overall system performance. Hybrid user association schemes, which overcome the drawbacks of both MMAP and MRSP-based user association schemes, may provide improved trade-offs between spectral and energy efficiencies.  
This paper has focused on exploiting the multi-user diversity considering low mobility users and perfect knowledge of CSI. Nevertheless, in presence of users with heterogeneous mobility, different scheduling criteria can be used for low and high-mobility users  \cite{mobility}  to achieve frequency as well as  multi-user diversity gains.

\bibliographystyle{IEEEtran}

\end{document}